\documentclass[sigconf]{acmart}
\AtBeginDocument{%
  }

\copyrightyear{2026}
\acmYear{2026}
\setcopyright{cc}
\setcctype{by}
\acmConference[CHI '26]{Proceedings of the 2026 CHI Conference on Human Factors in Computing Systems}{April 13--17, 2026}{Barcelona, Spain}
\acmBooktitle{Proceedings of the 2026 CHI Conference on Human Factors in Computing Systems (CHI '26), April 13--17, 2026, Barcelona, Spain}
\acmPrice{}
\acmDOI{10.1145/3772318.3791960}
\acmISBN{979-8-4007-2278-3/2026/04}

\usepackage{xcolor}
\usepackage{hyperref}
\usepackage{url}

\newcommand{\re}[1]{\textcolor{black}{#1}}

\begin{document}


\title{Exploring Collaboration Breakdowns Between Provider Teams and Patients in Post-Surgery Care}

%\author{}
%\affiliation{%
%  \institution{}
 % \city{}
  %\state{}
  %\country{}}
  %\email{}

\author{Bingsheng Yao}
\authornote{Both authors contributed equally to the paper.}
\author{Menglin Zhao}
\authornotemark[1]
\affiliation{%
  \institution{Northeastern University}
  \city{Boston}
  \state{Massachusetts}
  \country{USA}}
% \email{b.yao@northeastern.edu}

% \author{Menglin Zhao}
% \affiliation{%
%   \institution{Northeastern University}
%   \city{Boston}
%   \state{Massachusetts}
% \email{zhao.mengl@northeastern.edu}

\author{Zhan Zhang}
\affiliation{%
  \institution{Pace University}
  \city{New York City}
  \state{New York}
  \country{USA}}
% \email{}
% \email{zzhang@pace.edu}

\author{Pengqi Wang}
\affiliation{%
 \institution{The Ohio State University}
 \city{Columbus}
 \state{Ohio}
 \country{USA}}
% \email{}
 % \email{wang.19883@osu.edu}

 \author{Emma G Chester}
 \author{Changchang Yin}
\affiliation{%
 \institution{The Ohio State University Wexner Medical Center}
 \city{Columbus}
 \state{Ohio}
 \country{USA}}
 % \email{}
 % \email{Emma.Chester@osumc.edu}

% \author{Changchang Yin}
% \affiliation{%
%  \institution{The Ohio State University Wexner Medical Center}
%  \city{Columbus}
 % \email{}
 % \email{Changchang.Yin@osumc.edu}

 \author{Tianshi Li}
 \author{Varun Mishra}
 \author{Lace Padilla}
\affiliation{%
  \institution{Northeastern University}
  \city{Boston}
  \state{Massachusetts}
  \country{USA}}
  % \email{}
  % \email{tia.li@northeastern.edu}

% \author{Varun Mishra}
% \affiliation{%
%   \institution{Northeastern University}
%   \city{Boston}
%   \state{Massachusetts}
  % \email{}
  % \email{v.mishra@northeastern.edu}
  
% \author{Lace Padilla}
%  \affiliation{%
%   \institution{Northeastern University}
%   \city{Boston}
%   \state{Massachusetts}
  % \email{}
  % \email{l.padilla@northeastern.edu}

\author{Odysseas Chatzipanagiotou}
\author{Timothy Pawlik}
\affiliation{%
 \institution{The Ohio State University Wexner Medical Center}
 \city{Columbus}
 \state{Ohio}
 \country{USA}}
 % \email{}
% \email{OdysseasPanteleimon.Chatzipanagiotou@osumc.edu}

% \author{Timothy Pawlik}
% \affiliation{%
% \institution{The Ohio State University Wexner Medical Center}
%  \city{Columbus}
% \email{}
% \email{Tim.Pawlik@osumc.edu}

  \author{Ping Zhang}
\affiliation{%
 \institution{The Ohio State University}
 \city{Columbus}
 \state{Ohio}
 \country{USA}}
 % \email{}
 % \email{zhang.10631@osu.edu}

\author{Weidan Cao}
\authornotemark[2]
\affiliation{%
 \institution{The Ohio State University Wexner Medical Center}
 \city{Columbus}
 \state{Ohio}
 \country{USA}}
 % \email{weidan.cao@osumc.edu}

\author{Dakuo Wang}
\authornote{Corresponding author: weidan.cao@osumc.edu, d.wang@northeastern.edu}
\affiliation{%
  \institution{Northeastern University}
  \city{Boston}
  \state{Massachusetts}
  \country{USA}}
\renewcommand{\shortauthors}{Yao and Zhao et al.}

\begin{abstract}

% Managing post-surgery recovery at home after discharge requires ongoing collaborative effort among stakeholders in the provider teams and on the patients' side. 
% Despite prior work in HCI and CSCW identifying challenges in each stakeholder's work, such as patients' low literacy and compliance and providers' decision-making difficulty, there is a lack of a systematic investigation of collaboration breakdowns across the care pipeline.
% In the context of gastrointestinal (GI) surgery as a case study, we conducted semi-structured interviews with eight physicians, five nurses, and four patients to consolidate multi-stakeholders' perspectives on where collaborations fail in post-surgery care. 
% Two mutually reinforcing problem areas were identified.
% Within provider teams, over-complicated structures and weak inpatient–outpatient coordination slow information triage and yield inconsistent guidance.
% At the patient–provider boundary, clinic–home resource gaps and non-contextualized patient-generated data make clinical decision-making and patient self-care inaccessible.
% We outline design opportunities for intelligent systems that formalize task ownership and handoffs, contextualize co-temporal signals, and align care plans with home resources.

Post-surgery care involves ongoing collaboration between provider teams and patients, which starts from post-surgery hospitalization through home recovery after discharge. While prior HCI research has primarily examined patients’ challenges at home, less is known about how provider teams coordinate discharge preparation and care handoffs, and how breakdowns in communication and care pathways may affect patient recovery. To investigate this gap, we conducted semi-structured interviews with 13 healthcare providers and 4 patients in the context of gastrointestinal (GI) surgery. We found coordination boundaries between in- and out-patient teams, coupled with complex organizational structures within teams, impeded the “invisible work” of preparing patients’ home care plans and triaging patient information. For patients, these breakdowns resulted in inadequate preparation for home transition and fragmented self-collected data, both of which undermine timely clinical decision-making. Based on these findings, we outline design opportunities to formalize task ownership and handoffs, contextualize co-temporal signals, and align care plans with home resources.

\end{abstract}




% \keywords{collaboration, patient-provider communication, continuity of care, postoperative recovery, patient care, clinical coordination, invisible work, articulation work, clinical documentation, care transitions}


\begin{teaserfigure}
\centering
  \vspace{-.5em}
 \includegraphics[width=.93\textwidth]{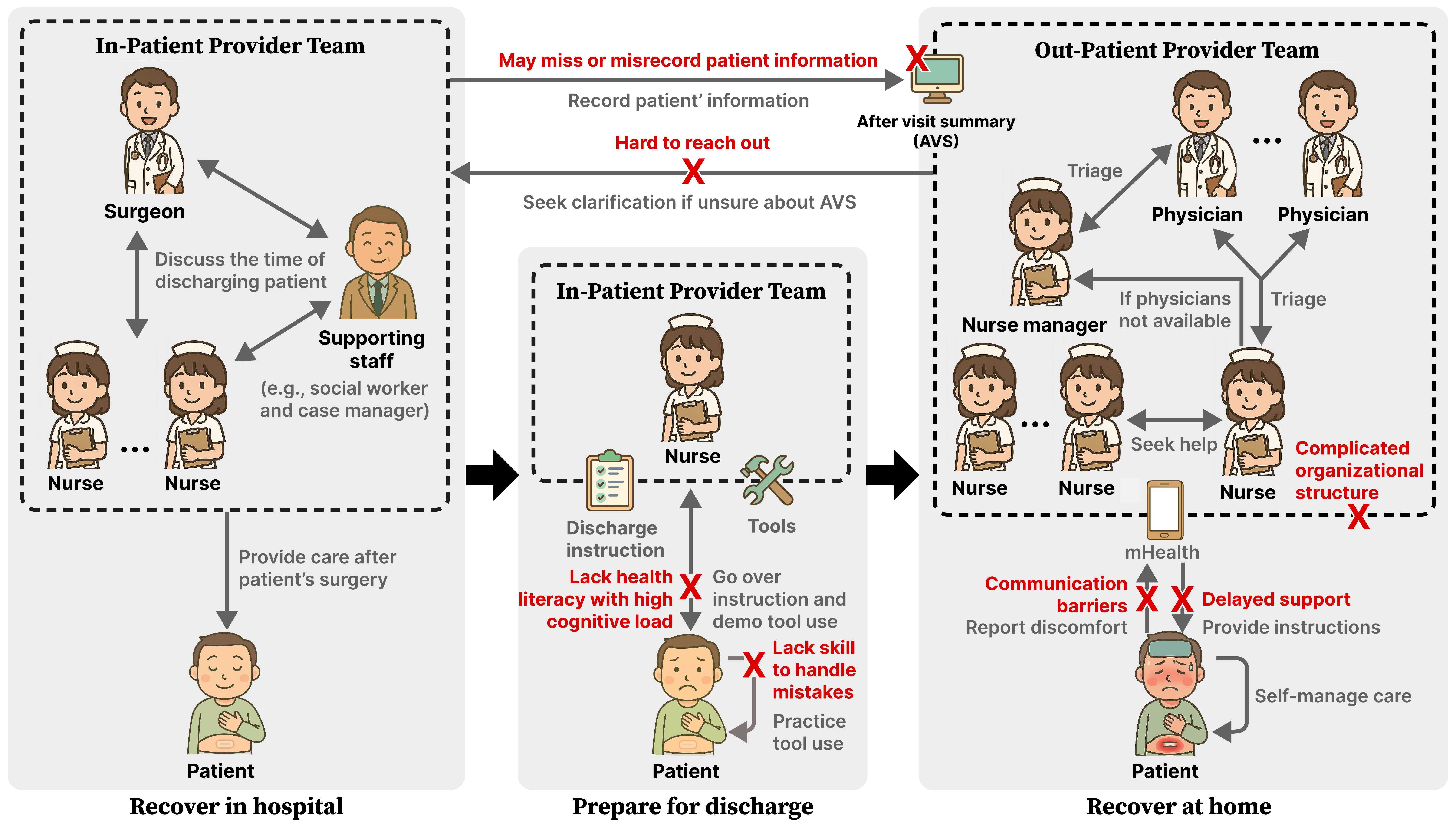}
 \vspace{-.5em}
  \caption{An illustration of how multi-stakeholders collaborate during the post-surgical care transition process. Through interviews with 13 care providers and four patients, we identified coordination and communication breakdowns between in-patient and out-patient care provider teams, and between patients and providers, which create challenges for patients to prepare self-management of care at home and can lead to a delay in care support.}
  \label{fig:teaser}
   \Description{Three-panel diagram showing the transition from hospital to home. Left panel, “In-Patient Provider Team”: a surgeon, two nurses, and a supporting staff member coordinate care and discuss discharge timing. A patient icon below indicates recovery in the hospital. Middle panel, “Prepare for discharge”: a nurse teaches discharge instructions and demonstrates tools; the patient practices. Callouts mark barriers: limited health literacy, high cognitive load, and lack of skills to handle mistakes. A connector labeled “After Visit Summary (AVS)” carries information to the right panel and back to the inpatient team for clarification. Red warnings note risks: missed or misrecorded information and difficulty reaching the inpatient team. Right panel, “Out-Patient Provider Team”: two physicians, a nurse manager, and several nurses triage messages in a complex structure. At home the patient reports symptoms via mHealth and receives guidance. Red X icons indicate communication barriers and delayed support. Bold arrows show the flow: recover in hospital → prepare for discharge → recover at home.}
\end{teaserfigure}

% \begin{figure*}[htbp]
%   \includegraphics[draft=false,width=\textwidth]{Figures/workflow.pdf}
%   \caption{In this paper, we conducted semi-structured interviews with 13 healthcare providers and four to systematically examine their workflow of conducting SICs and challenges they face during their workflow. We also explored the AI opportunities and concerns in supporting ED-specific SIC workflow with our participants. Finally, we derived design guidelines grounded in our findings.}
%   \label{fig:study_procedure}
%    \Description{}
% \end{figure*}

\maketitle

\section{Introduction}

Patients' recovery journey after major surgeries often constitutes a critical care transition. 
Immediately after surgery, patients, while hospitalized, rely on a tightly coupled infrastructure where surgeons, nurses, and social workers collaborate to provide continuous monitoring and immediate intervention~\cite{Reddy2002Finger,bardram2000temporal, thompson2017hospital, stephens2015postoperative}. 
Once discharged from the hospital, patients must rely on themselves or family caregivers to manage their post-surgery care at home. 
Nevertheless, complicated post-surgery care routines, which often include medication schedules, wound hygiene, and symptom monitoring, are mentally and physically demanding for patients and caregivers, who do not possess care expertise~\cite{de2010patients, pollack2016closing,zghab2024s, mccaughan2018patients}.
Thus, patients could feel unprepared for wound care and are often afraid to remove their surgical dressings, while family members worry about accidentally damaging the wound~\cite{tobiano2023patient}.
Such a care transition is particularly demanding following Gastrointestinal (GI) surgery because patients face a high risk of severe complications, such as organ failure~\cite{Mazzotta2020PostoperativeIA,thomas2020complications,ellis2021defining}, while simultaneously being required to manage mobility impairments and specialized medical apparatuses~\cite{yang2025recover}, such as surgical drains and feeding tubes~\cite{durai2009use, sethuraman2017tube, shrikhande2013post}.

% These non-experts must immediately manage intricate routines, including medication schedules, wound hygiene, and symptom monitoring, without confidence~\cite{pollack2016closing,zghab2024s,Kripalani2007DeficitsIC, mccaughan2018patients}. 
% For example, a study showed that patients generally feel unprepared for wound care and are often afraid to remove their dressings, while family members worry about accidentally damaging the wound~\cite{tobiano2023patient}.
% The transition from hospital to home is particularly demanding following Gastrointestinal (GI) surgery. 
% %%revise
% In this context, patients face a high risk of severe complications, such as organ failure~\cite{Mazzotta2020PostoperativeIA,thomas2020complications,ellis2021defining}, while simultaneously managing mobility impairments and specialized medical apparatuses like surgical drains and feeding tubes~\cite{Zhang2020EvaluationOP}.
% Even though they have family members around, their family often lacks confidence and worries about accidentally damaging the wound~\cite{tobiano2023patient}.
%%revise

Various assistive technologies have been designed by the HCI and medical informatics communities to support patients' post-surgery at home. 
For instance, personal informatics, wearables, and AI-based remote patient monitoring (RPM) systems were designed to allow providers to track patients' physical well-being~\cite{klasnja2010blowing, oh2022patients,raj2017pgd,zhang2022predicting,keys2024think}, assist patients' daily self-care at home~\cite{yang2025recover, klasnja2010blowing}, and enable early identification of abnormal signals about complications~\cite{bignami2025wearable, van2023early}.
Nevertheless, patient adherence to care procedures and the effectiveness of self-care management with these technologies remain unsatisfactory, largely because patients lack the necessary resources, health literacy, or self-efficacy to use these tools confidently and correctly in their daily care routine~\cite{pollack2016closing,mishra2016not}.
Recently, several studies suggest that a largely overlooked reason underlying these home-based struggles lies upstream, specifically, during the discharge preparation stage~\cite{kripalani2007promoting,hesselink2012improving,kattel2020information,kim2022discharge}.
Poorly executed care transitions from hospital to home can lead to adverse outcomes, such as incorrect treatments, medication errors, and even mortality~\cite{leithaus2025health}.
For patients who are not able to effectively learn and practice home care procedures or align their expectations with their care team before leaving the hospital, their capacity for self-care at home and the support from providers are often compromised~\cite{pollack2016closing,kangovi2014challenges, yang2024wish}. 
However, the limited cognitive load inherent in the fast-paced hospital environment, particularly following a major surgery, often impedes a patient's ability and cognitive readiness to learn critical tasks for home-based care~\cite{chugh2009better, acher2015using, yang2025recover, pollack2016closing}.

We position \textbf{the discharge preparation stage as a complex socio-technical coordination problem} critical to subsequent long-term care quality at home, instead of as a simple informational transaction in the discharge summary.
In particular, establishing the patients' readiness for post-discharge care is an outcome achieved through collaborative efforts involving multiple provider teams working with the patient~\cite{pollack2016closing} during the preparation stage.
High-quality discharge preparation requires the alignment of shared artifacts (e.g., discharge summary)~\cite{bardram2000temporal,fitzpatrick2013review,star1989institutional}, established routines~\cite{Feldman2000OrganizationalRA}, and temporal alignment across different clinical roles~\cite{reddy2001coordinating,bardram2000temporal}.
For example, prior work shows that whether a patient can be discharged on time depends on whether nurses, physicians, and case managers stay aligned around the same plan~\cite{bardram2000temporal}.
Despite the criticality, the specific socio-technical mechanisms that drive coordination among provider teams and patients during the discharge preparation stage remain largely underexplored.
Specifically, there is a lack of a grounded understanding of how provider teams coordinate their workflows and share information critical to the transition of care, how provider teams work together with patients to co-construct patients' readiness for self-care at home, and what coordination breakdowns may occur during the preparation stage.
Moreover, the current community lacks understanding of how the effectiveness of discharge preparation influences patients' self-care at home after discharge, and what design considerations could emerge to support the continuity of care during and after the transition of care from hospital to home for surgery patients.

To address these gaps, this study investigates the specific socio-technical mechanisms during the discharge preparation stage in post-GI surgery as a case. We selected post-GI surgery because the patients face a significant risk of life-threatening complications while managing complex self-care, such as drain and wound management.
Our work is guided by three research questions: \\
\indent \textbf{RQ1:} \textit{How do provider teams coordinate the workflow and information sharing required to prepare GI surgery patients for discharge? } \\
\indent \textbf{RQ2:} \textit{How do these provider teams interact with patients to align expectations and co-construct readiness for the transition home?} \\
\indent \textbf{RQ3:} \textit{What design considerations should inform socio-technical systems that bridge the gap between hospital-based discharge preparation and home-based recovery?}

% \begin{enumerate}
%     \item \textbf{RQ1:} How do provider teams coordinate the workflow and information sharing required to prepare GI surgery patients for discharge? 
%     \item \textbf{RQ2:} How do these provider teams interact with patients to align expectations and co-construct readiness for the transition home?
%     \item \textbf{RQ3:} What design considerations should inform socio-technical systems that bridge the gap between hospital-based discharge preparation and home-based recovery?
% \end{enumerate}

We conducted a qualitative study involving semi-structured interviews with both GI providers (e.g., surgeons, nurses, nurse managers) (N=13) and GI surgery patients (N=4).
By probing into perspectives from both groups, we connect the provider-side work of discharge preparation with the patient-side experience of transitioning home and managing self-care.
Our findings identified breakdowns in provider-provider coordination that create fragmented and conflicting surgical patients' documentation and unclear responsibilities (as shown in Fig.~\ref{fig:teaser}). These coordination breakdowns between provider teams during discharge preparation directly translate into inadequate patient preparation, compromising both patient home-based care and the ability of providers to intervene effectively after discharge.
In addition, we identified design considerations for sociotechnical systems to support patients' continuity care from hospital-based discharge preparation and home-based recovery.

\section{Related Work}

\subsection{Collaboration Between Provider Teams and Patients In Post-Surgery Care}

In the U.S., patients often get discharged within days after the surgery, and enter a period of heightened vulnerability to life-threatening complications (e.g., wound infection, sepsis, and organ failure)~\cite{bruce2001measurement, semple2015using, mezei1999return}. 
These complications pose significant threats to patient survival and quality of life \cite{wang2025risk}. 
For instance, research indicates that 72.7\% of patients develop surgical site infection (SSI)~\cite{yao2021research}, which extends hospital stays and the likelihood of multiple reoperations~\cite{sorensen2005risk}. 
Unlike the hospital setting, where patients receive professional post-surgery care, at-home recovery shifts much of the management of care to patients themselves, with many lacking adequate health literacy to recognize complication-related signals and properly manage their care, such as wound care~\cite{jakobsson2023everyday, andrus2002health}. 
Therefore, managing post-surgical recovery cannot be achieved by patients acting alone; rather, it fundamentally depends on ongoing collaboration between patients and their provider teams, who must work together to monitor recovery, recognize early warning signs (e.g., wound redness or swelling, severe pain, and fever~\cite{andersson2010patients}), and intervene before conditions worsen.

Researchers identified multiple challenges in post-surgery care after patients return home.
First, physical distance between patients and providers makes it difficult for patients to obtain timely support from providers~\cite{de2010patients}, and for providers to monitor patients’ recovery progress.
Even when patients seek help by calling providers, provider availability is often limited, so early warning signs can go unnoticed.
Digital tools (e.g., MyChart~\cite{mychart}) offer a channel for patients to share updates, but their limited usability excludes those with lower digital literacy~\cite{vanderhout2025patient}.
Although patients receive instructions for self-managing care, these are often generic and fail to address their unique needs~\cite{yang2024wish}.
Furthermore, the inconsistent guidance and ambiguous escalation thresholds can contribute to poor recovery outcomes~\cite{zghab2024s}.

Given the substantial challenges patients face when transitioning to self-management at home, HCI researchers have investigated the underlying factors contributing to these difficulties. 
Through interviews and observations with patients during hospitalization and post-discharge, Pollack et al. identified that successful care transitions depend on three critical elements: knowledge, resources, and self-efficacy. 
They found that discontinuities at both system and individual levels undermine patients' ability to safely conduct self-care~\cite{pollack2016closing}. 
However, while extensive research has focused on patient challenges during home recovery, the preparatory stage at discharge, when care responsibility transfers from clinical teams to patients, remains significantly underexamined.

\subsection{Socio-Technical Breakdowns and Interventions During Care Transitions}
Prior work on care transitions has consistently shown that the post-discharge period is clinically fragile and organizationally fragmented, with reviews documenting high rates of post-discharge errors, suboptimal handovers, and delayed or incomplete discharge summaries between hospital and primary care clinicians~\cite{kripalani2007promoting,hesselink2012improving,kattel2020information}. Systematic reviews of computer-enabled discharge tools and broader handover interventions suggest that electronic summaries, structured planning, and shared follow-up can improve timeliness, legibility, and completeness of information, but their effects on outcomes such as readmissions and mortality remain mixed and methodologically hard to compare~\cite{motamedi2011efficacy,hesselink2012improving,kattel2020information}. Qualitative and population-specific studies reveal how these information problems are embedded in messy, cross-setting workflows: transfers between nursing homes and emergency departments are described as labor-intensive, loosely coordinated, and driven more by institutional routines than residents' clinical needs, while hospital-to-skilled-nursing-facility transitions for people living with dementia are distressing, behaviorally complex, and hampered by systematic under-communication of dementia-related information~\cite{mccloskey2011qualitative,gilmore2017transitions}.

Similar patterns appear in transitions from hospital to home health, where communication between hospital and home care providers for older adults is sparse, largely one-way, and characterized by limited trust, unclear contact pathways, and a striking lack of empirical guidance on how to improve discharge communication~\cite{kim2022discharge}. Complementing these setting-specific accounts, work on specialty–primary care coordination shows that well-designed documents alone are insufficient without shared expectations, role clarity, and ongoing relationships, underscoring that transitions are socio-technical accomplishments shaped by artefacts, digital systems, and interprofessional collaboration~\cite{kim2015care,kripalani2007promoting,mccloskey2011qualitative,gilmore2017transitions,kim2022discharge}.

While prior research has documented the vulnerabilities of post-discharge care and the fragmentation of cross-institutional communication, it has largely framed 'discharge preparation' as a transactional information handoff. This perspective overlooks how multidisciplinary teams and patients collaboratively build the capacity for home-based self-care during the inpatient stay. Consequently, we lack a systematic understanding of how care teams coordinate workflows and information sharing, and how they interact with patients to establish realistic expectations and capabilities for home recovery.
These gaps build the foundation for our RQ1 and RQ2.

\subsection{The Usage of Remote Patient Monitoring and AI Technologies in Post-Surgery Care}

Remote Patient Monitoring (RPM) is becoming important to supporting home recovery, which supports healthcare providers to track patients' recovery progress and intervene when needed~\cite{tabacof2021remote, farias2020remote, mclean2024implementation}. 
Some researchers in the HCI community have explored the potential use of RPM for different patients, such as post-surgery cancer patients~\cite{timinis2025co} and youth concussion patients~\cite{yao2025more}.  
Mobile monitoring improves adherence, quality of life, and reduces readmissions~\cite{dawes2021mobile}, which extends supervision beyond clinic hours. 
RPM tools span passive wearables~\cite{staehelin2024patient}, and smart wound-monitoring systems~\cite{keegan2023implementation}.
Some devices, including wearables, can also support patients in self-management~\cite{keys2024think, james2025integrating}.
However, health systems can diverge from patients' at-home realities, which idealize patients' behaviors clash with their constraints~\cite{burgess2022care}. 
Moreover, many assumed that patients have enough motivation and digital literacy, but overlooked the potential information overload, limited access, and communication gaps~\cite{walton2022patients, brajcich2021barriers}. 
Therefore, the shift from passive inpatient care to self-management at home can be difficult~\cite{zghab2024s}.

Recent research in HCI and the medical informatics community has focused on integrating AI to support both clinical workflows and patient recovery. For example, combining wearable data with electronic health records allows for the early identification of complications and readmission risks~\cite{wu2025cardioai}. 
Some researchers explored using AI for identifying wound deterioration based on the wound image~\cite{anisuzzaman2022image}.
Moreover, Large Language Models (LLMs) are increasingly used to collect patients' data through conversations, contextualize patient-generated data for clinical dashboards~\cite{yang2025recover, dwyer2023use, hao2024advancing, li2024beyond, montagna2023data, ni2017mandy, wei2024leveraging}, and provide active, empathetic follow-up~\cite{yang2024talk2care, ferrara2024large} as well as clinical prediction~\cite{kruse2025large, xu2024mental}.

Successfully deploying these technologies requires careful integration into clinical practice. 
Research emphasized that AI systems should augment, not replace, clinical deliberation and must be designed to handle uncertainty and accountability~\cite{yang2019unremarkable, cai2019helloai, zajkac2023clinician, burgess2023healthcare}. 
Besides, systems are evolving to support collaborative sensemaking and intermediate reasoning, such as using counterfactual explanations to reduce overreliance~\cite{lee2023counterfactual} and creating feedback loops between experts and models~\cite{lee2021rehab, zhang2024rethinking}. 
Clinical decision support systems integrate symptom trajectories and physiological trends to fosterunderstanding of patients~\cite{wu2025cardioai}, with recent studies suggesting LLM-based decision systems can function as team members to support evaluation of decisions and improve usability and trust~\cite{rajashekar2024human}.

Although these technologies are useful in some clinical settings, there is limited research on design opportunities that support the socio-technical systems bridging discharge preparation and home-based recovery. This gap is especially urgent for patients who are fragile with a high risk of life-threatening complications after surgery, yet have to manage their own care at home. Therefore, the gap motivates our RQ3.

\begin{table*}[!t]
  \caption{Demographics of participants in semi-structured interview. 
  P refers to patients, and CP refers to healthcare providers.}
  \label{tab:demographics}
  \begin{tabular}{cclcc} 
    \toprule
    Provider ID\# & Gender & Job Title & Department & Year of Practice \\ 
    \midrule
    CP1 & Female & Nurse Practitioner & Surgical Oncology Department & 7 years \\ 
    CP2 & Female & GI Surgeon & Surgical Department & 6 years \\ 
    CP3 & Male & GI Surgeon & Surgical Department & 15 years \\ 
    CP4 & Female & Nurse Practitioner & Surgical Department & 10 years \\ 
    CP5 & Female & Nurse Practitioner & Surgical Department & 12 years \\
    CP6 & Male & GI Surgeon & Surgical Oncology Department & 11 years \\ 
    CP7 & Female & GI Surgeon & Surgical Oncology Department & 8 years \\ 
    CP8 & Female & Nurse Practitioner & Surgical Department  & 9 years \\ 
    CP9 & Female & Director of Operations & Surgical Department  & 32 years \\ 
    CP10 & Male & GI Surgeon & Surgical Department & 16 years \\
    CP11 & Male & GI Surgeon & Surgical Department & 18 years \\ 
    CP12 & Female & Nurse Practitioner & Surgical Department & 20 years \\ 
    CP13 & Female & Registered Nurse & Surgical Department & 14 years \\ 
    \midrule
    Patient ID\# & Gender &  & Type of GI Disease  & Age Range\\
    \midrule
    P1 & Female &  & Pancreatic Cancer & 50-59 \\ 
    P2 & Male &  & GI Stromal Tumor & 50-59 \\
    P3 & Female &  & GI Sarcoma Tumor & 60-69 \\
    P4 & Female &  & Colon and Liver Cancer & 60-69 \\
    \bottomrule
  \end{tabular}
\end{table*}

\section{Semi-Structured Interviews}
To investigate the breakdowns among multi-stakeholders during the hospital-to-home transition, we conducted a qualitative study using semi-structured interviews with two primary stakeholders in this process: 13 healthcare providers who specialize in GI surgery and four patients who had recently undergone these procedures. 
Our research design was guided by the need to uncover coordination work among clinical teams and connect it to the lived experience of patients managing their recovery at home.
This dual perspective gave us a grounded understanding of breakdowns that could occur across different provider teams and patients during discharge preparation and helped identify design opportunities to better support this care transition.

% To address our research questions, we conducted a two-stage study. 
% First, we interviewed 13 healthcare providers who specialize in GI surgery to understand their challenges and needs when collaborating with patients in post-operative care, especially after patient discharge following a major GI procedure. Moreover, we examine the opportunity of AI and RPM technologies to support patient-health care team members during their collaboration. 
% We also conducted interviews with four patients who had GI surgery to gain their insight about challenges and needs working with healthcare providers around post-operative care, especially when at home. The findings from these interviews helped inform several proposed evidence-based design guidelines using AI and RPM technologies to support patient-provider collaboration in post-operative care, especially in the home setting. In this section, we present details related to participants (both patients and healthcare providers) and the procedure of the study.

\subsection{Research Participants}
We employed a combination of purposive and convenience sampling to recruit participants.
\textbf{13 GI providers} were recruited from the professional networks of our team's clinical experts.
Our purposive sampling was critical for capturing the multiple perspectives of a care provider team that consists of multiple roles and stakeholders. 
Clinical participants are from multiple hospitals within the same large academic medical system. 
Some worked in general acute-care hospitals, while others worked in hospitals specialized in oncology, including GI cancer care. 
We recruited surgeons (N=4), inpatient nurses (N=3), outpatient clinical nurses (N=4), and nurse managers (N=2).
This variety provided a multi-perspective view of the collaborative work involved during transition.

We also recruited \textbf{four patients} who had undergone a major GI surgery within the previous six months. A central argument of our paper is that provider-side coordination failures directly impact patient outcomes.
Therefore, our study prioritizes a deep investigation of the provider-side work practices, which are currently under-examined. The patient interviews were designed to complement this focus by capturing the causative effects of the providers' work and the interaction between patients and providers during the discharge preparation stage in the subsequent patient care when the patient is at home.
Table~\ref{tab:demographics} provides a detailed overview of participant demographics, their roles, and their clinical experience. 
The study was conducted in the U.S., and all participants were living in the U.S. at the time. 
The study was reviewed and approved by the Institutional Review Board at the first author’s institution, and all participants provided informed consent.

\subsection{Study Procedure}

We conducted semi-structured interviews via Zoom, with each session lasting approximately 30 to 45 minutes to accommodate the demanding schedules of providers.
All interviews were audio-recorded and transcribed for post-study analysis.
We designed two interview protocols for care providers and patients.
For \textbf{care providers}, we began by asking providers to recall a recent experience when they had a patient who developed post-operative complications after GI surgery.
By grounding the discussion in this concrete case, we can ask follow-up questions about team composition and coordination, communication breakdowns, and challenges of remote monitoring after the patient went home.
We then probed deeper into the creation and use of artifacts like discharge summaries and the difficulties of cross-team communication.
Although these technologies are useful in clinical settings~\cite{yao2025more, yang2025recover}, little work examines how to support fragile, high-risk patients as they transition home and must manage their own care. Thus, our interview protocol included questions to serve as a probe to explore the applicability of existing technology proposals to continuity-of-care work in post-GI surgery to better answer RQ3.

For \textbf{patients}, we centered the interview on their experience of transitioning home and after discharge. 
Each session with patients lasted about 60 to 90 minutes.
We ask follow-up questions primarily about challenges and needs related to following discharge instructions, managing their daily care, dealing with symptoms, seeking help from providers, and collaborating with providers and their caregivers.
To explore needs and design opportunities around technologies, such as wearables and CAs, we used three storyboard scenarios as conversational probes to help patients better understand the functions of these technologies.
Each scenario in the storyboard, corresponding to one technology, was designed based on insights from prior studies~\cite{yang2024talk2care, wu2025cardioai, yang2025recover}. 
The first storyboard illustrated a voice-based assistant that checks on the patient’s health daily and allows patients to verbally report symptoms and self-care activities. The second storyboard depicted wearable sensors that continuously track physiological indicators and trigger alerts. The third storyboard presented a mobile health app for sending wound images, messaging providers, and scheduling appointments. 
The storyboard’s goal was to support patients' understanding of these innovative technologies, rather than to validate the storyboard's scenarios. Moreover, we discussed with them any of their concerns about these technologies.

\subsection{Data Analysis}

To protect privacy and align with the IRB, we reminded all participants at the beginning of each interview not to disclose any personally identifiable information, such as names of care providers, patients, and specific institutions. All interviews were audio-recorded and transcribed verbatim with participants’ prior consent.
We then analyzed the transcripts adhering to the six phases of reflexive thematic analysis (RTA)~\cite{braun2021thematic}, treating analysis as an ongoing, interpretive, and researcher-involved process. Before formal coding, the research team read through all transcripts multiple times to become familiar with the data and wrote initial analytic memos to capture early impressions. 
We then generated inductive, open-ended codes, attending to segments that spoke to participants’ experiences. We coded for both semantic (explicit) content and latent (underlying) concepts. Codes were not derived from a fixed a priori coding frame but evolved through collapsing, splitting, and renaming. 
The team met regularly to review codes and cluster them into candidate subthemes. We utilized thematic maps to visually organize these clusters and explore relationships between varying codes. 
Through reflexive discussions on how our positionalities shaped the data, we iteratively refined the thematic maps. We moved back and forth between the full dataset and the emergent structure, checking if themes and subthemes coherently captured patterned meanings. We revised boundaries and definitions until we established a final set of themes. 
Finally, we synthesized the findings. We compared and contrasted themes across patient and provider participants to identify shared patterns and salient differences; these comparisons structure the findings reported in this paper.

\section{Findings}

Our analysis of interviews with GI surgery providers and patients reveals a series of breakdowns in the hospital-to-home care transition. These breakdowns are not isolated incidents but are rooted in systemic issues of coordination, communication, and a fundamental disconnect between the well-monitored and maintained environment of the hospital and the complex realities of patients' homes. 
We present our findings in three parts. 
First, we \re{described the coordination of workflows and information sharing among GI providers, the breakdowns during the discharge preparation (Section~\ref{provider_coordination}, which answered RQ1.
Second, addressing RQ2, we examined the interactions between GI providers and patients as they align expectations, co-construct readiness for the transition home, as well as the unexpected challenges in self-care caused by unreadiness for discharge~\ref{patient_provider_coconstruction}.
Lastly, we identified design considerations, including both opportunities and concerns, for sociotechnical systems to support hospital-based discharge preparation as well as home-based recovery (Section~\ref{design_consideration}) to answer RQ3.}
Figure \ref{fig:teaser} illustrates the current clinical workflow for providing care to post-GI surgery patients across hospital and home settings, as well as the main points of breakdown.

% \begin{figure*}[htbp]
%   \includegraphics[draft=false,width=\textwidth]{Figures/workflow.pdf}
%   \caption{In this paper, we conducted semi-structured interviews with 13 healthcare providers to systematically examine their workflow of conducting SICs and challenges they face during their workflow. We also explored the AI opportunities and concerns in supporting ED-specific SIC workflow with our participants. Finally, we derived design guidelines grounded in our findings.}
%   \label{fig:study_procedure}
%    \Description{}
% \end{figure*}

\subsection{The Coordination Between Different Provider Teams and Information Sharing for Preparing GI Patients for Discharge}
\label{provider_coordination}

\subsubsection{Discharge Artifacts That Fail as Boundary Objects Undermine Cross-Team Coordination}

Supporting post-GI surgery recovery requires coordination across a fragmented specialty chain: surgeons and inpatient nurses manage the acute operative phase, while a distinct team of outpatient nurses and wound care specialists must interpret that care plan weeks later without having witnessed the surgery.
In the inpatient setting, discharge planning for GI patients is particularly dense, requiring surgeons to dictate precise protocols for drains, ostomy appliances, and dietary restrictions into the EMR, details that are critical for survival but easily mistranscribed.

Importantly, patient discharge instruction is a patient-facing document, whereas the after-visit summary and consult notes are clinician-facing.
Participants emphasized that, in practice, outpatient nurses often use the discharge instruction to go through the important items that patients need to keep in mind when managing care at home. As CP9 explained, the discharge instruction functions as ``\textit{kind of like the source of truth for getting a patient ready to go home}''.
As for the after-visit summary, it includes information, such as patients' surgical-related information and recovery plan, acts as a boundary object~\cite{star1989institutional} (i.e., shared artifacts help people cooperate without consensus), documented in the EHR for other providers to quickly understand patients' health conditions when they need to provide patient care.

\begin{quote}
\textit{``Nurse, practitioners or physician assistants are the ones who write the discharge summaries... That note, then gets co-signed by the attending, who perform the surgery.''} (CP5)
\end{quote}

However, outpatient nurses reported that these documents are often flawed, containing inaccuracies, omissions, or contradictions that disrupt their workflow.
Outpatient nurses reported that errors in the summary, such as listing a drain as "removed" when it is still present, or citing incorrect medication dosages, create immediate safety risks when they attempt to execute a care plan based on flawed data.
For example, they often observe conflicting information in the patient visit summary. 
Because outpatient providers were not involved in the inpatient care process, nor in creating these documents, they find it very difficult to validate the correct information.
As a result, outpatient providers often make significant articulation work, spending time trying to contact the inpatient team for clarification.
However, the reconciliation is often unsuccessful.
The surgeons are often extremely busy with surgeries and hard to reach at times, and even if they get a chance to meet, the surgeons often forget the details of the patient's medical history, including the surgical date and the type of surgery, because patients and providers may not have met for months.
% (I think this quote can be put together with the prior errors or missing information part)
\begin{quote}
\textit{``...but sometimes, you know, you might find some errors on the avs (the after visit summary), or maybe the patient says, Oh, I'm not on that medication anymore, you know.''} (CP9)
\end{quote}

These breakdowns point to a need for documentation integrity checks and cross-team retrieval that surface contradictions early.
Inpatient and outpatient sub-teams are structurally separated but have to coordinate to provide post-surgery patient care.
Reaching back across teams is often impractical: when outpatient staff ``\textit{call into the hospital and ask to speak to the resident on service... the problem is... that resident isn't working today}'' (CP9). Even within the same clinic, layered triage can slow escalation of patients' information, as CP3 mentioned that ``\textit{usually the calls go to an RN... and then the nurse will make a decision about... what things they can comfortably take care of}''.

\begin{quote}
\textit{``So as an outpatient clinical nurse, basically, what we did was like triage phone calls or messages that came into the inbox... We would triage those messages, and we would send those on in like an S Bar format to the provider.''} (CP13)
\end{quote}

During this coordination, the patient visit summary, often expected to be a boundary object, may have inaccurate or lack information to carry a shared understanding about patients' health conditions. 
The flaws lead to outpatient nurses being forced into articulation work (e.g., spending a lot of time to confirm the accuracy of information) to reconcile inconsistencies. 
However, it can be hard to reconcile inconsistencies when outpatient nurses are urgently needed. 
Outpatient teams are responsible for providing proper care based on the patient's visit summary.
However, they often lack reliable ways to verify the accuracy of the decisions (e.g., meditation use) on the patient visit summary. 
Based on this finding, we move beyond simply identifying these errors to analyze their specific nature in GI surgery, such as listing a drain as "removed" when it is still present or citing incorrect dosages, and their systemic causes, which stem from surgeons dictating dense protocols into rigid EMR templates that are easily mistranscribed. Crucially, we demonstrate the consequence casuing by the failed boundary objects, forcing outpatient nurses into invisible "articulation work" to reconcile data and creating immediate safety risks.

\begin{quote}
\textit{``I always want somebody to double-check me. So if I have, if I can have technology double-check me, then. I think that would be an easy workflow if it was there to in just from a nursing perspective. Okay, here's my, here's my abs. Let me have technology double-check me. Are there any things that it caught? Okay, looks good to me. Let's let me move on, whatever, I mean, I feel like that could be worked into a workflow.''} (CP9)
\end{quote}

%%%

\subsubsection{Resource Scarcity and Layered Triage Slow Provider Coordination and Disrupt Post-Surgery Care}

The aforementioned challenges are compounded by systemic issues within healthcare organizations that further impede timely response to complications.
A recurring friction in GI recovery stems from the incompatibility between the high-touch requirements of managing abdominal wounds and ostomies, and the low-touch, resource-scarce structure of standard outpatient follow-up.
Many institutions lack the personnel to conduct proactive remote monitoring of patients after discharge, such as a follow-up phone call, since personnel are "a very expensive resource", as CP2 mentioned. 
Unlike specialized or well-resourced centers—such as cancer institutes that can afford routine follow-up calls, a few providers (CP2, CP7, CP11) noted that many general hospitals must rely on patients to self-report symptoms or initiate contact when complications arise.
As a result, this reactive approach creates delayed feedback loops, placing dual responsibility on patients to both recognize problems and successfully navigate the clinic's communication channels.

\begin{quote}
\textit{``So their (providers) ability is very limited frankly, to really take care of these patients or spend a lot of time with these patients, because they are spread thin.''} (CP10)
\end{quote}

Even when patients do reach out, the process of triaging information to the appropriate provider proves slow and uncertain due to two structural barriers. 
First, general triage nurses frequently lack the specific wound or drain care training required to distinguish between normal post-operative inflammation and signs of deep organ space infection, necessitating escalation.
However, identifying the appropriate provider contact is complicated by the fact that many providers are already occupied with in-clinic responsibilities. 
Second, even when nurses know whom to escalate to, communication must pass through multiple intermediaries. 
For instance, a nurse may forward a concern to a specialist's nurse, who then relays it to the specialist when available.  
While this layered system is designed to manage workload distribution, it actually produces redundancy and prolongs the time before patients receive expert attention.
Sometimes, patients do not even know when or if they will receive a response from their providers.

\begin{quote}
\textit{``I don't recall or see any notes where they (providers) give a timeframe on when that would be responded to [the patients]. ''} (CP2)
\end{quote}

These structural dynamics become particularly evident in wound monitoring practices. 
Providers (P2, P6, P10) emphasized that wound appearance is a critical marker of recovery and potential complications. 
To support remote monitoring, some patients attempt to send photographs through patient portals like MyChart. 
However, the visual complexity of GI surgical sites renders patient-generated photography largely illegible without clinical lighting or precise angles.
Since nurses cannot reliably interpret these photographs, they must forward them to physicians.
In many cases, physicians ultimately determine that the images remain inconclusive and instruct patients to return to the clinic for in-person evaluation. 
This cycle creates a double inefficiency: surgeons must waste time deciphering ambiguous images of abdominal sites, only to eventually default to an in-person visit to palpate the wound or assess depth—assessments that 2D images cannot capture.

\begin{quote}
\textit{``Yeah, like too close, or you, too close, really grainy, couldn't tell whether the redness at the incision site was actually like a darker pink or red. really hard to tell [what's the condition of the wound]. ''} (CP4)
\end{quote}

This case reveals how distributed clinical work becomes undermined by the intersection of resource scarcity and organizational redundancy. Prior CSCW studies of healthcare have examined coordination breakdowns in high-resource environments, often focusing on intra-professional communication or information systems integration. 
The GI post-surgery context surfaces a more visceral dimension: how fragile communication infrastructures fracture under the weight of complex, fast-changing physiological recoveries that require visual and tactile assessment rather than just data reporting.
Our analysis shows that resource constraints and organizational layering combine to create significant inefficiencies, where each handoff compounds delays instead of fostering coordination. 
This extends CSCW understandings of distributed care by foregrounding the structural and infrastructural barriers that arise when specialized knowledge must be mobilized across fragmented teams without reliable, proactive monitoring systems.

\subsection{Patient-Provider Interaction for Aligning Expectations and Co-Construct Readiness for the Transition Home}
\label{patient_provider_coconstruction}

During discharge preparation, GI providers rely heavily on a single, dense instructional encounter. CP2 and CP9 explained that nurses meet with patients and their family members to walk them through the discharge instruction document. In this session, nurses highlight the key tasks patients must manage at home—such as wound care, drain management, and pain control—and clarify symptoms like fever, vomiting, or new drainage that may require contacting the care team. When patients raise questions or note inconsistencies (e.g., medication lists that do not match their current regimen), nurses address them while reviewing the paperwork (CP9).
This one-time explanation serves as the primary educational moment before patients leave the hospital, shaping their understanding of what to expect and how to manage recovery at home. For high-risk cases in which patients are unlikely to manage complex tasks such as tube care or dressing changes independently, providers arrange for visiting nurses or skilled nursing facilities to bridge the gap.

\begin{quote}
\textit{``The nurse will actually print out the discharge instructions. Then they highlight and circle like the different key areas, and they go over it with the patient and their family while they're there in the hospital. ''} (CP2)
\end{quote}

\subsubsection{The Disconnect Between Passive Discharge Preparation and the Active Management Affected by the Lived Reality of Home}

% add patients' perspectives.
The transition from hospital to home represents a profound challenge for most GI surgery patients.
This transition is defined by a dramatic shift in responsibility: while providers control and monitor all aspects of care inside the hospital, patients are suddenly tasked with maintaining clinical sterility for complex abdominal drain care within their own unadapted living spaces.
A central feature of many GI surgery recoveries is the use of surgical drains to prevent fluid buildup and enable close monitoring of healing. 

During hospitalization, these devices are managed exclusively by clinical staff. However, before discharge, the burden of care rapidly transfers to patients and their families. 
Nurses and physicians attempt to bridge this gap with hands-on instruction, requiring patients to demonstrate competence before discharge.
As CP1 put it, ``\textit{they (patients) have to do it in the hospital before they can go home, kind of like an evaluation, that they are doing it correctly.}'' 
Providers actively encourage family participation during these sessions, recognizing that greater exposure increases the chances of success at home. 
The expectation from care providers is clear: patients have to replicate the same drain care procedures, such as squeezing and pulling along the drain tube to clear blockages, without support.

However, GI patients often failed to translate into reliable execution once stripped of the clinical infrastructure and oversight available on the ward.
GI providers described situations where patients, despite demonstrating competence at discharge, failed to maintain proper drain care after leaving the structured environment of the hospital. Some patients never change the drain until the next clinical visit, as CP1 described: ``\textit{even though we're instructing them (patients). Sometimes they come back into the office 2 weeks later, and they've never changed the dressing, and it's soaking wet.}'' The gap suggests just-in-time, symptom-aware guidance that adapts procedures to home constraints.

Beyond drain care, many patients are also asked to perform self-measurement tasks at home, such as tracking food intake, recording bowel movements, monitoring blood pressure, checking vital signs, or weighing themselves. Compared to the technically demanding routines of drain care, these forms of self-monitoring might appear simpler. 
However, in the context of GI recovery, even these seemingly simple recording tasks proved physically demanding, as the logistics of gathering data conflicted with the patients' limited mobility and need for rest. 
As PP1 noted, recording and interpreting health data felt burdensome, especially when combined with post-operative discomfort, limited household resources, or a home environment that is poorly suited to recovery. 
The hospital’s design, which places supplies within arm's reach, does not carry over to the home setting, creating a context where the simple act of retrieving a logbook becomes a painful physical hurdle due to the GI surgery.
As a result, the transition exposes a gap: what seems straightforward under supervision can break down quickly under real-world conditions, leading to adverse events such as infections or emergency readmissions.

\begin{quote}
\textit{``Well, I mean, you know, the challenges are, you know, unless you can have all that right beside your chair, or your bed, or, you know, so… you gotta record that stuff, like, morning, noon, and night. So you constantly happen to move the stuff with you, or… or go to, you know, whatever room it’s in to do it all. And like I said, if you’re not feeling good, and you’re not… you know, getting up and down is very painful at first. You just don’t want to get up and go do that.''} (PP1)
\end{quote}

Our analysis identified that self-care is obstructed by the specific somatic reality of post-GI surgery, where severe core muscle soreness physically prevents the movements required to inspect wounds or manage drains.
Three patient participants described how sore muscles made it difficult to move, get out of bed, or even reposition themselves, yet post-surgery care frequently demands mobility for activities such as wound inspection, drain maintenance, and self-measurement. The expectation to perform these tasks often conflicts with the lived experience of acute postoperative pain, fatigue, and limited mobility. 
Despite this, provider training during discharge rarely addresses how to manage care in the presence of these symptoms. 
This oversight means that patients are often unprepared for the reality of managing their care while symptomatic. For example, while still in pain, a patient may interpret discomfort as a signal to rest, which can discourage them from performing tasks such as changing dressings or recording vital signs.

Discharge education typically focuses on the mechanics of the procedure, neglecting the contextual strategies needed to perform these tasks while managing acute abdominal pain and restricted torso mobility due to GI surgery.
The lack of guidance on how to adapt care routines when experiencing symptoms reduces patients’ confidence and increases their anxiety about self-management. When PP2 was told at discharge that the next clinic visit would be in a month, he was worried about handling problems on his own and was not ready for home self-care.
In contrast, PP1 had previous surgical experience and took steps to prepare her home environment before discharge. For example, she installed a grab bar in the bathroom and modified her bed, which increased her confidence and made it easier to manage post-discharge routines.

\begin{quote}
\textit{``When I was discharged, it was like, we'll see you in a month. The surgeon, and the surgeon was, like, super high confident, like, yep, we'll see you in a month. And I remember at the time, I was like, that seems like a long time from now... You know, there felt like there needed to be some things to check inter, you know, in between the time, and meeting back with them.''} (PP2)
\end{quote}

The inadequacy of preparation leads to a large proportion of patients ending up requiring unplanned home health nursing, skilled nursing facilities, or other forms of professional support, especially when the complexity or frequency of care exceeds what families can manage. 
Previous studies have reported that discharge instructions often lack personalized information to support effective patient care at home. Our findings indicate that addressing challenges in post-discharge care requires more than revising written instructions.  
Discharge preparation for self-care is vital to closing the gap between hospital and home.

\subsubsection{The Importance of Early Caregiver Involvement in Preparing for Discharge Readiness and the Influence of Relational Dynamics on Patient Recovery}

While caregivers were not interviewed due to the study’s scope, both patients and GI providers described family members as essential informal caregivers who must engage before discharge. Providers noted that time pressure limits how much tailored preparation they can offer, which leads to instructions that emphasize broad precautions rather than guidance grounded in the patient’s living context.
Many patients return home with severely limited mobility, which makes tasks that seem simple (e.g., moving devices or walking to record vital signs) difficult or impossible. The challenge is inseparable from the material constraints of the home, where restricted mobility and limited physical support infrastructure complicate even the most routine activities.

To address these challenges, early caregiver involvement is essential. Multiple clinicians emphasized that integrating caregivers into the educational process earlier significantly improves outcomes. As P6 noted, “We like to have the
family members also present. I think it helps to have more people hear it, the usually the more successful it is”. 
Furthermore, feedback from family members regarding their own capacity to provide care directly influences clinical decisions. 
Caregivers often identify limitations in their ability to assist during the discharge planning phase. P6 observed that “sometimes the family members can tell you immediately, like, I won’t be able to do this,” referring to tasks such as wound care. This admission forces the medical team to arrange for external support, such as home health aides or skilled nursing facilities, prior to the patient leaving the hospital. In many cases, this external support is a rigid requirement for discharge, with P6 estimating that “probably 50\% of the time, if not more.”

In some cases, caregiver involvement begins even before surgery. PP1 described how she and her husband prepared for her transition in advance. As her primary caregiver, he joined pre-surgery discussions with physicians, who explained that her mobility and self-care abilities would be compromised after discharge. Together, they developed an adaptation plan and installed assistive devices at home to support bathing and daily routines before she underwent GI surgery. As a result, her return home was relatively smooth because the domestic environment had already been adapted to her needs.
This positive deviation shows how early, active caregiver participation can eliminate barriers that many other patients later face. It also shifts the temporal focus of transition work. While prior literature often situates support and resource management during or after the transition~\cite{pollack2016closing}, this case demonstrates that the pre-surgical phase is a critical window for discharge preparation. 

\begin{quote}
\textit{``My husband and I are pretty good at being prepared for recovery, so we have, like, the walker, the handles, like, by the toilet, we had already redone our bathroom and have handles in the shower, along with a seat, so we were all prepared for that, but it all came in very handy...So that was definitely helpful to have all that in place before I came home.''} (PP1)
\end{quote}

However, while caregivers play a structural role in recovery, the quality of the patient–caregiver relationship can profoundly shape home care. PP4 noted that although his wife was supportive, his relationship with his 19-year-old daughter became openly hostile during his post-surgical and chemotherapy recovery. She blamed him for “bringing cancer on himself,” yelled at him for ordinary behaviors like chewing or breathing too loudly, and created an environment where, as he said, “it would be a lot easier if my daughter wasn’t there.” Instead of providing comfort or help, the family dynamic turned shared spaces into stressors and pushed him toward isolation.
Although he considered himself “psychologically very strong,” he still said he was “pretty much by myself,” given his wife’s ~80-hour workweeks and his daughter’s antagonism. This case illustrates that caregiver–patient conflict does not just affect mood; it determines whether the home can function as a supportive recovery setting at all.
Together, these findings complicate prior HCI work that often implicitly treats family members as supportive components of the care infrastructure~\cite{mynatt2001digital, foong2020you}, and may overlook the influence of overtly unsupportive family members during patients' recovery process.

%fail to align expectation and its nageative consequence caused by unreadiness for
%the transition 

\subsubsection{Misaligned Symptom Understanding and Digital Barriers Reduce the Discharge Readiness}
\label{unaligned_expectation}

GI providers reported that complications after a GI surgery are their common and significant concern.
These complications can manifest through behavioral signals (e.g., inability to tolerate food), physiological signals (e.g., fever), and wound-related signals (e.g., swelling or bleeding).
To reduce the risk of developing complications, CP11 emphasized the importance of early collection of these signals.
To collect these signals after patients' discharge, hospitals often rely on follow-up calls.

However, providers noted that these calls are resource-intensive and not always feasible. Even when nurses do make calls, it can be difficult to assess a patient’s real condition.
Our analysis identifies two primary mechanisms driving this difficulty. 
The first is the failure to align expectations between patients and GI providers regarding self-care standards after a GI surgery. 
For example, CP13 explained that when a patient claims they are drinking "enough," she needs to verify the exact amount because the patient's expectation of what constitutes "enough" often fails to align with clinical goals.
Moreover, CP2 reported that it can be difficult to accurately assess a patient’s surgical wound condition based solely on patients' verbal descriptions, as she noted that ``\textit{patients are unreliable at describing their wounds}''. 
Even if patients can take pictures of the wound and upload them to the patient protocol, the quality of the image is often not good enough for assessment.
Consequently, the burden of identifying abnormal signals shifts heavily to patients, who are expected to initiate contact through calls or patient portals.

\begin{quote}
\textit{``The timing makes a huge difference. It makes a difference from when you can intervene and help them and potentially avoid another admission or an emergency room visit.''} (CP11)
\end{quote}

For better support of patients' self-care at home, some GI providers provide their numbers and online medical protocols to patients for reaching out.
However, patients still face several challenges that hinder their ability to report health concerns promptly. 
First, patients are often unaware of their inability to distinguish between routine post-operative symptoms and abnormal complications.
As CP11 noted, patients might dismiss significant post-operative pain as "normal," which can delay necessary intervention.
This finding is substantiated by patient accounts, which reveal a fundamental difficulty in calibrating the severity of their own symptoms. On one hand, patients often normalize potential warning signs.
For instance, PP4 viewed severe pain simply as an inevitable part of the post-surgical experience and focused solely on requesting pain medication rather than questioning the source of the discomfort. 
Conversely, PP2 experienced anxiety over benign healing processes, unable to determine if minor redness around the wound indicated a routine occurrence or a serious complication. 
This finding aligned with what CP11 mentioned that ``\textit{patients may experience that (pain) even outside of the recovery period of an operation, so they might not think it's of significance, when, in fact, it might be of significance after the surgery}''.
Furthermore, when GI providers explicitly align expectations regarding physical restrictions, such as heavy lifting, patients often struggle to adhere to these boundaries in practice. We observed instances where patients, despite knowing the prohibition on lifting, inadvertently exerted themselves, which led to wound rupture and subsequent hospital readmission.

\begin{quote}
\textit{``Patients may experience that (pain) even outside of the recovery period of an operation, so they might not think it's of significance, when, in fact, it might be of significance after the surgery.''} (CP11)
\end{quote}

Even when patients recognize an abnormal signal and want to reach out to their GI providers, technology friction can block reporting.
As CP2 mentioned, many patients, especially older adults, struggle to start using hospital portals (e.g., MyChart), with difficulty remembering passwords to log in, which makes digital reporting effectively inaccessible.
Moreover, PP3 and PP4 mentioned that they were reluctant to use mobile applications.
This reluctance did not originate upon discharge; rather, it reflected a long-standing lifestyle preference.
PP3 explicitly stated that ``\textit{If you're talking to my demographic of 60-plus 60 years old plus. We might not be so into using apps....I would love to get rid of my phone, period}''.

Despite these established habits, the post-discharge protocol relies on phone-based portals for symptom checking, such as sending a wound image.
At the moment, patients were physically exhausted and mobility-impaired; they had to adopt interaction patterns they disliked. 
Consequently, routine digital tasks can compound their distress, converting simple interface navigation into a significant cognitive burden during their most vulnerable recovery phase.
When health resources are locked behind cumbersome interfaces, GI patients can be excluded from the care they expect.
These barriers call for low-friction symptom capture and channel access that work for surgical patients.
\begin{quote}
\textit{``A lot of times, they want things simple... Some of our patients are older and don't want to do anything that's digital.''} (CP12)
\end{quote}

%Lastly, generic discharge instructions, even with some personalization, may still fail to account for individual patient needs and home contexts.
%For example, a common restriction for post-GI surgery patients is weight lifting. 
%However, due to the limitation of discharge instructions, patients may lack awareness of the maximum weight that they can lift, which can cause severe pain and require re-hospitalization. 

These findings exposed a critical gap in collaborative alignment between provider expectations and patient realities at home. 
Specifically, the study highlights a novel dimension of situational impairment. The specific temporal window where physical exhaustion compels patients to revert to established habits (e.g., avoiding technology) directly clashes with discharge protocols that demand the adoption of new digital behaviors. Consequently, insufficient preparation is not just a lack of medical education, but a failure to design care infrastructures that respect the 'bandwidth constraints' of recovering bodies and the established digital boundaries of post GI surgery patients, especially for older adults.

\subsection{Design Opportunities for Sociotechnical Systems to Bridge the Gap Between Hospital-Based Discharge Preparation and Home-Based Recovery}
\label{design_consideration}

\subsubsection{Need For Context-Aware and Coordinated Technologies to Capture Real-Time Contextual Signals to Support Post-Surgery Care}

Our interviews revealed that providers share patients' concerns about post-discharge care, specifically regarding the high risk of complications that could lead to readmission.
Due to the high risk of complications, several providers (CP2, CP6, CP7, CP11) expressed a strong interest in AI-based tools for risk prediction. 
However, the mere presence of an AI risk score for the high-risk patientsis insufficient, which aligns with the previous HCI studies~\cite{zhang2024rethinking, cabitza2024never, wu2025cardioai}.
One participant (CP10) described an existing AI-powered readmission risk score that is available to them at discharge. 
This tool has seen little adoption because it falls short of clinical needs.
The primary criticism is its opacity; the single score can not lead to any actions.
Providers argue that prediction features must highlight key symptoms contributing to complications.

Furthermore, an effective prediction feature may also highlight the key symptoms or factors contributing to the complications. 
For instance, if a model predicts a high risk of dehydration, it should also identify the correlated symptoms and suggest targeted interventions, such as a revised fluid intake plan.
This granular information could support care providers in their critical clinical judgments at discharge.
On the one hand, care providers need to ensure a patient is in a stable health condition and capable of managing their care at home before discharge. 
On the other hand, there are institutional pressures to manage bed capacity and reduce the average length of stay.
As CP13 stated, patient turnover is very fast; patients can be discharged just one or two days after surgery to meet the goal of reducing patients' length of stay in the hospital.
Therefore, detailed and actionable AI features can be a crucial resource for care providers to justify decisions and advocate for the patient's safety.
\begin{quote}
\textit{``Unless you tell me what I can do to minimize the risk of admission. I'm going to send this patient home because I'm being judged both on readmission rates and also on length of stay. So you can't really just because I'm getting that information that they are higher risk for readmission. I can't just keep them in the hospital.''} (CP10)
\end{quote}

However, current AI support technologies are not yet available at discharge, and patients continue to face multiple challenges in managing their care at home.
To bridge this gap, participants increasingly advocate for home monitoring through wearables and Conversational agents (CAs). They view wearables as optimal for capturing physiological parameters, including heart rate, temperature (CP7), and sleep patterns (CP6). 
Meanwhile, CAs can systematically collect structured symptom reports (e.g., pain, eating, nausea) while helping differentiate normal variations from concerning signals.
Nevertheless, participants emphasized concerns about data overload. 
Simultaneously, participants highlighted the burden of excessive data presentation. CP2 and CP6 specifically advocated for systems to display only abnormal findings rather than comprehensive data streams to minimize time investment and cognitive overhead. 
Furthermore, when device anomalies appear in isolation without cross-device integration, clinicians lack sufficient contextual information to assess whether patients are experiencing genuine complications or benign variations.
As CP3 explained, elevated heart rate exemplifies this ambiguity. It could indicate pain (e.g., suggesting complications) or simply normal activity. Therefore, monitoring systems must coordinate across devices and capture contextual information when anomalies emerge. 
For instance, when wearables detect heart rate spikes, a CA could immediately ask the patient about concurrent activity, pain, dizziness, and anxiety levels.
Presenting an anomaly alongside this time-aligned context transforms an ambiguous data point into a meaningful insight. This integrated approach is necessary to move beyond simple data collection and toward systems that effectively support clinical sensemaking.

\begin{quote}
\textit{``So kind of joining heart rate data with some of those types of things might give you a better idea of their hydration status and their nutrition and things like that.''} (CP3)
\end{quote}

%reorganize
To better support care providers in double-checking the summary, CP9 noted that AI could help care teams by automatically double-checking the accuracy of patient visit summaries, thereby reducing workload while improving reliability. 
This stands in contrast to much of prior HCI research, which has traditionally emphasized human validation of AI outputs~\cite{yao2025more, zhao2025designing}.

\subsubsection{Bridging the Home-Clinic Gap Through Low-Burden Sensing and Pragmatic Voice Interaction}
Building on the patient participants' unawareness and uncertainty about abnormal symptoms, described in Section~\ref{unaligned_expectation}, participants emphasized the need for active, real-time monitoring to differentiate normal recovery from complications. 
Specifically, P2 identified wearables as a primary mechanism to track physiological indicators, including temperature, pulse, sleep patterns, breathing patterns, and heart rate. 
These metrics are critical in post-GI surgery care as they often serve as early warning signs for severe GI complications, such as internal bleeding. 
Moreover, P1 located the primary value of wearables in their capacity for passive data collection. 
By automating measurement, these devices eliminate the need for physical movement, which is a frequent source of post-surgical pain. Given this heightened sensitivity, P1 framed physical comfort as a non-negotiable design constraint. 
P2 echoed this, arguing that because patients already grapple with significant somatic discomfort, the technology must not introduce secondary physical irritation. 
She specifically contrasted adhesive patches with rigid commercial smartwatches, favoring the former. The flexible form factor of a patch aligns with the physical reality of recovery, whereas rigid hardware becomes intrusive during prolonged rest.

\begin{quote}
\textit{``Well, because that (wearables) is kind of an automatic thing, you don't have to, like, really stop and do that. It kind of does it for you, so it's an automatic thing... they want you to record things, like your body temperature, your weight, your blood pressures, your sugar.... when you're not feeling good, it's just not something you want to do...getting up and down is very painful at first. You just don't want to get up and go do that.'} (PP1)
\end{quote}

While wearables have the ability to monitor physiological metrics and capture any abnormality, it is hard for them to capture the condition of surgical wounds.
As we mentioned in Section 4.1, patients' symptom reports need to be along with their wound image to allow GI providers to comprehensively assess the wound condition.
However, due to the fact that patients' self-report image quality is low, P1 noted that it would be beneficial if a technology could actively assess image quality at the point of capture. 
By providing real-time guidance on lighting and focus, the technology could ensure GI providers receive clinically useful data without waiting for patients to repeatedly retake photos. 
This capability would create a more efficient feedback loop: it allows patients to receive faster reassurance without guessing, and it enables providers to identify wound deterioration early enough to facilitate timely intervention.
This finding extends prior CSCW and HCI work on patient-generated health data by showing that, in post-surgical care, symptom reports and wound images cannot function as independent data streams; GI providers rely on them together to form a clinically actionable assessment.

Although a technology-driven image-taking is promising, there remains a friction point. 
Mobile applications require a level of physical capability that post-surgical patients often lack. 
During the early phases of GI recovery, patients are frequently bedridden, which renders interaction with screens or paper contact lists impractical. 
In response to the accessibility challenge, CAs emerged as a viable solution. 
PP2 emphasized that voice-based interaction allows patients or those without caregivers to report symptoms and ask for help without physical exertion. 
Beyond physical accessibility, CAs could also support patients to reach out to their GI providers through a complex organizational structure. 
As detailed in Section 4.2, the hospital's organizational structure is complex, and patients find it hard to reach out to their GI providers.
Therefore, patients wish a CA could intelligently route these requests, directly and automatically connecting patients to the appropriate personnel and abstracting the complex clinical hierarchy into a single voice interface.
Collectively, P1 emphasized that these technologies offer a critical safety net for patients with limited support from caregivers.

\begin{quote}
\textit{``If you are in bed, you're in pain, you don't have access to a phone, or you don't want to‚ be typing on a small screen... this would be the way to do it, they're laying in bed, they're in pain, they they can reach out [to a CA], you know, verbally.''} (PP2)
\end{quote}

\subsubsection{Concerns of Technology in Post-GI Surgical Home Care}
Despite the functional promise of CAs and wearables, participants raised concerns about the interaction itself. The ephemerality of voice-only interfaces created cognitive anxiety. P4, a visual learner, said “pure voice” felt insecure because he could not verify what he heard; with a speaker, he feared missing critical details while “the words are being processed.” Unlike text, which he could “review… a few times mentally,” audio is transient. He directly linked this fear of missing information to “anxiety,” indicating that CAs in high-stakes health contexts need a visual log to complement voice.

Participants also rejected CAs that mimic human social norms, diverging from prior work that emphasizes human-like personalities. P4 noted that although AI voices are technically fluent, they lack the therapeutic quality of “being listened to,” and the interaction feels “limited” because the system cannot ask relevant follow-up questions like a human caregiver. P2 expressed frustration with CAs that are “overly polite” or “solicitous,” arguing that pleasantries such as “I hope you’re having a good day” obstruct utility. She preferred the system to “cut through all the pretend nice human stuff” and simply provide information, explaining that because the agent “isn’t a nice human, it’s a machine,” adopting a personality creates friction rather than rapport.

Beyond interaction style, participants wanted clear control over when care is initiated and how data is shared. P3 drew a sharp line between user-driven and system-driven interactions. She was comfortable issuing voice commands but rejected proactive systems, saying she would not tolerate a device that “suddenly wakes me up one morning” to check on her—contrasting with work on proactive health monitoring~\cite{yang2025recover, yang2024talk2care}.
This desire for agency extended to wearables. Although automatic reporting is possible, P3 insisted that data must not be sent automatically. She wanted to review it herself first to decide “whether to contact a nurse or not,” maintaining her role as the primary decision-maker.

\section{Discussion}

% \tianshi{I think the discussion will be stronger if you can engage with your actual findings (e.g., specific themes emerged from the qualitative coding) more frequently and in more depth. Use the findings as the support evidence of the points you make. One tip I personally like is to directly refer to the related results section in the discussion (e.g., Section x.x)}

\subsection{Collaboration Breakdowns and Overlooked Technological Support During the Transition from Provider-Led to Patient-Led Post-Surgery Care}

During the transition from hospital to home after GI surgery, the responsibility for recovery shifts dramatically, from being managed by the clinical team to being managed almost entirely by patients themselves. However, as our findings show, this shift is not adequately supported. Hospital staff tend to focus on immediate clinical care, while systematic investment in personalized discharge preparation and patient training is lacking. This oversight can leave patients poorly prepared to manage their care at home, resulting in higher risks of complications and unnecessary readmissions.

Our analysis shows that although GI surgery recovery is complex, current discharge preparation rarely addresses the real challenges patients face at home. Complex provider structures, time pressure, limited resources, and a narrow focus on acute clinical tasks undermine the delivery of comprehensive, context-sensitive self-management training. As a result, patients are expected to identify complications, use digital portals, and perform technically demanding tasks such as drain care without sufficient skills, confidence, or support. This creates a dangerous implementation gap: patients may misinterpret symptoms, struggle with digital tools, or rely on generic instructions that fail to address individual vulnerabilities. And even when they seek help, limited resources and organizational barriers frequently delay timely intervention.

Another structural barrier is the inaccuracy of handover documents, such as discharge summaries. However, we move beyond simply identifying these errors to analyze their specific nature in GI surgery, such as listing a drain as "removed" when it is present or citing incorrect dosages, and their systemic causes, which stem from surgeons dictating dense protocols into rigid EMR templates that are mistranscribed. Crucially, we demonstrate that these errors function as failed boundary objects, forcing outpatient nurses into "articulation work" to reconcile data and creating immediate safety risks when care plans are executed based on flawed information.

Making matters worse, our study found that coordination between inpatient and outpatient teams is often undermined by inconsistent documentation and a lack of shared understanding. Handover documents such as discharge summaries were frequently incomplete, inaccurate, or ambiguous, forcing outpatient nurses to spend significant invisible work verifying information, tracking down unavailable colleagues, or making decisions based on unreliable data. This fragmentation further undermines a safe care transition and shifts the burden of interpretation onto patients who are already underprepared.

Current patient–provider collaboration in HCI largely focuses on interactions during hospitalization or after discharge~\cite{schroeder2017supporting, seo2021learning, murphy2017ambiguous}, such as patient-generated data as boundary-negotiating artifacts~\cite{chung2016boundary}, collaborative care-plan creation~\cite{zhao2021supporting}, and shared health tracking in the hospital~\cite{mishra2018supporting}. 
However, our findings show that the most critical breakdowns occur earlier, during the preparation phase, when responsibility is being transferred. Inadequate preparation shifts the burden into the home, making later breakdowns downstream effects of insufficient discharge readiness. Both patients and providers often overlook this phase, missing chances to build practical skills and contextual knowledge. When preparation is rushed or generic, even advanced technologies cannot compensate for gaps in patients’ basic understanding or self-efficacy.

This misalignment suggests that both practice and research should shift some of their focus back to proactive, patient-centered, and context-sensitive preparation at the point of discharge. Only through hands-on training, scenario-based education, and clearly defined support channels before patients go home can we truly empower patients and families to manage independently, reduce complications, and prevent avoidable readmissions.

\subsection{Infrastructural Torque, Coordination Amplification, and Resource-Coordination Paradox in Post-GI Surgery Care}
Our findings align with CSCW and HCI traditions that view healthcare as sustained by sociotechnical infrastructures rather than isolated tools. Prior research shows how information systems embed themselves in organized work, shaping visibility and redistributing responsibility~\cite{bowker2000sorting,star1996steps}. Infrastructure theory further highlights how technologies and organizations form interdependent arrangements requiring ongoing maintenance~\cite{hanseth2008theorizing,lee2006human}. Moreover, studies reveal how patients perform articulation work to repair fragmented systems~\cite{gui2019making,star1999layers}.
We situate our study within this lineage but extend the lens to the institutional boundaries of post-surgery care. Coordinating GI care is not simply information transfer but an infrastructural problem requiring alignment across disconnected units, artifacts, and clinical routines. One of the important findings is that the discharge instructions often function as brittle boundary objects that fail to bridge inpatient and outpatient settings. We revealed the infrastructural torque that forces GI providers to bend the care trajectories to fit a rigid documentation process rather than having systems that fully support the care itself.

Moreover, our analysis extended prior work on organizational dependencies by modeling how structural complexity creates new bottlenecks and shapes care delivery. While infrastructure studies emphasized that work relies on layered human arrangements~\cite{star1996steps,lee2006human}, we identified a specific pattern in GI post-surgery care: “coordination amplification.” Each additional role or triage layer meant to distribute workload instead creates new handoffs that amplify the delay of care.
Specifically, we identified a dependency chain in which inpatient surgeons rely on case managers to translate plans into discharge documents, while outpatient nurses must track down available physicians and rely on their specific expertise to respond to patient portal messages. 
Consequently, this complexity converts additional personnel and tools into increased articulation work, not increased care capacity.

Based on the identified structural complexity, we applied articulation work~\cite{schmidt1992taking} (i.e., a term that describes the compensatory efforts required to bridge gaps across roles, systems, and organizational boundaries) to a new and important post-surgery care setting.
We identified how GI providers performed articulation work without support.
In our study, GI provider participants mentioned that they have to call multiple care providers to reconcile conflicting documentation in the EHR. 
As a result, GI providers fall into recursive verification loops that produce predictable delays. 
The challenge points to a need for detecting discrepancies and adding rationale annotations. 
Being able to resolve conflicts during AVS creation and carry that rationale forward would reduce the articulation work, save clinical resources, and facilitate patient care.

Finally, these coordination failures require rethinking how technological interventions are introduced. Although HCI and health informatics often see automation or remote monitoring as a solution to extend clinical reach~\cite{liu2025scaffolded}, our analysis reveals a core tension: adding new services or technologies may create additional organizational layers and queues. For instance, when patients contact providers through mHealth tools, nurses often consult other specialists before replying, which produces secondary delays.
We describe this as a “resource-coordination paradox”: if new tools do not improve coordination mechanisms, their benefits can be absorbed by the extra coordination they require. Technological interventions must therefore consider the underlying infrastructure and how multiple stakeholders may use the tool, so that the advantages of new resources are not erased by rising coordination costs.

\subsection{Designing Collaborative Device Ecosystems to Capture Correlated Clinical Signals to Support Clinical Assessment of GI Complications}
Recent HCI research has introduced a range of tools for capturing and generating patient-produced health data. Much of this work focuses on using wearables to support self-care management~\cite{zhang2022predicting, keys2024think, tadas2023using, hardcastle2020fitbit} and to enable shared decision-making where patients and providers review these data together~\cite{chung2016boundary, mentis2017crafting}. 
Moreover, scholars have explored data fusion techniques that combine streams from different sensors (e.g., environmental versus physiological temperature).
Yet wearables cannot capture subjective experiences such as pain intensity—an especially important indicator of complications in GI post-surgery care. To address this gap, conversational agents have been used to gather narrative symptom reports through natural language~\cite{liu2025scaffolded, yang2024talk2care, yang2025recover}, complementing objective wearable data. Recognizing this potential, recent work attempted to integrate these heterogeneous streams into a unified clinician dashboard~\cite{wu2025cardioai}.

However, our findings reveal two critical design gaps in these existing approaches.
First, there is a lack of an active "coordination" mechanism between the devices to collect data meaningfully. 
For example, if a wearable detects a fever, we should consider whether CAs or mHealth tools should be triggered to prompt the patient to photograph their wound.
Even when multiple data types are available, they often remain visually siloed rather than integrated into a view that supports clinical synthesis, which undermines real-world usefulness. We revealed that valid judgment in high-risk post-surgical care may depend on detecting synchronous events, such as a temperature spike occurring alongside increased pain. When interfaces fail to make these cross-tool, cross-temporal links explicit, providers must perform “invisible work”~\cite{star1999layers} by mentally integrating scattered data to form a hypothesis, which raises cognitive load and risks missing critical intervention windows.

Finally, we argue that distinct data-collection devices must operate as a “collaborative whole.” Tools should be linked through explicit handoff mechanisms, shared data models, and context-sensitive triggers. Ideally, an anomaly detected by a wearable would automatically trigger a CA or MyChart inquiry to collect immediate context, such as new symptoms or a wound photo, producing a more complete and temporally aligned picture of the patient’s condition. Future work should examine how to design these coordinated workflows so that each component contributes to a unified clinical decision pathway.
At the same time, system design must account for the physical and cognitive limitations of post-surgery patients. When technologies are built in silos, the burden of coordination is shifted onto patients, who must navigate inconsistent interfaces, manage multiple accounts, and repeat data entry. Future research, therefore, needs to identify the limits of patient acceptance and determine how to integrate these tools without compounding the physiological and psychological strain of recovery.

\subsection{Designing for Physical Frailty and Continuity of Care in Post-GI Surgery Care} 
\label{for_patients}
Our study shows that continuity of care after GI surgery is a fragile chain spanning pre-surgical preparation, discharge readiness, and home self-care. We extend prior work on home setup during rehabilitation~\cite{axelrod2009reality} by showing that proactive home preparation (e.g., grab bars) is a critical socio-technical component of discharge for high-risk surgeries. Without this preparation, even advanced monitoring cannot compensate for managing recovery in an under-prepared environment. Technologies such as RPM should therefore support anticipatory care work by helping patients understand their home constraints and risks before surgery.
Once discharged, patients are expected to self-measure and generate data, yet prior work already shows challenges with PGHD—data overload and articulation work~\cite{west2018common}, and uncertainty about what providers want~\cite{oh2022patients}. In our context, the primary barrier is physical: many patients are too frail or in too much pain to reach devices, and some find wearables themselves uncomfortable. Physical comfort must therefore be a core design constraint for post-surgery technologies.

Finally, our study shows that a central challenge for patients at home is determining whether symptoms are normal and when signs such as wound redness warrant action. Prior work notes that patients need contextualized feedback to judge health changes~\cite{grimme2024my}.
Although participants suggested using wearables to detect abnormalities, raw numerical data rarely helps them decide whether urgent care is needed. 
Further, self-tracking data only becomes meaningful when transformed into boundary objects that can facilitate meaningful information flow through communication between patients and providers~\cite{chung2016boundary}. We argue that technical support must therefore go beyond collecting vital signs and guide patients in generating multimodal, clinically relevant data. 
For example, when a patient reports wound redness, a CA guides them to take a clinically useful wound photo. Together, these insights shift health-data interpretation from single-modality readings to a multimodal, surgery-specific, and preparation-sensitive way to continuous care.

\section{Limitations}

Our study has several limitations. First, while our primary focus was on the under-examined provider-side coordination, and patient interviews served to contextualize those findings, a larger patient sample would offer a more comprehensive view of the patient experience. 
% This study had several limitations. The sample included only four patient participants, which limited the generalizability of our findings. While the small sample size could restrict the diversity of perspectives captured, we reached thematic saturation, as no new themes emerged in the final interviews. 
We also relied on semi-structured interviews to surface participants’ challenges, needs, and desires. While interviews provided in-depth insights into participants’ values and perceptions, they are limited in capturing participants’ actual behaviors and longitudinal experiences. 
Alternative methods, such as observational fieldwork or diary studies, could complement interviews by revealing how participants’ needs and challenges evolve over time and in real-world contexts.

Although the current study focused on GI surgical procedures, there are many distinct types and care pathways within surgery. While the diversity of cases in the current sample allowed us to surface prominent cross-cutting challenges and needs, we did not conduct a granular analysis of differences among different complex surgical procedures or patient trajectories. This topic remains an important direction for future work. Finally, our study focused on understanding multiple stakeholders’ needs and the opportunities for innovative technologies in post-surgical GI care; however, we did not implement and evaluate the system leveraging these innovative technologies to address these needs. Future research can develop and evaluate such systems to assess their usability, additional needs, and potential limitations.

\section{Conclusion}

In this study, we conducted a semi-structured interview study with both GI surgery patients and their care provider teams to identify collaboration breakdowns in post-surgery patient care. 
Our findings reveal how fragile coordination within and between clinical teams leads to extensive articulation work and downstream challenges for patients. We show that flawed boundary objects, such as the discharge summary, and a significant contextual gap between in-hospital training and the realities of at-home care, undermine patient readiness and safety.
This paper investigated the collaborative breakdowns in the transition from hospital to home following GI surgery through semi-structured interviews with various roles in the care provider team and patients.
The challenges identified are not the result of individual failings but of systemic issues where resource constraints and fragmented organizational structures impede effective care. 
The work contributes a grounded understanding of the complex transition and points toward specific design opportunities. 
Future systems can better support the care network by structuring the patient handoff, providing provider teams with contextual, time-aligned data, and tailoring care plans to the constraints of the home setting.

\begin{acks}

This work was funded in part by the National Institutes of Health under award numbers R01AI188576 and R01CA301579, and by the Northeastern University Tier-1 Research Grant. The content is solely the responsibility of the authors and does not necessarily represent the official views of the National Institutes of Health.

\end{acks}

\bibliographystyle{ACM-Reference-Format}
\bibliography{sample-base}

\end{document}